%% file: lmcp3_preprint.tex
\documentclass[iop]{emulateapj}

\newcommand{\degrees}{\mbox{$^{\circ}$}} 
\def\sqig{$\sim$}
\def\lsi{LS\,I\,+61\degrees 303}
\def\ls5039{LS\,5039}

\def\cxou{CXOU\,J053600.0-673507}
\def\J1018{1FGL J1018.6-5856}
\def\dem{DEM L241}
\def\kms{km\,s$^{-1}$}
\def\deg{\degrees}
\def\msun{$M_{\odot}$}
\def\ergs{erg~s$^{-1}$}
\def\ergcm2s{erg\,cm$^{-2}$\,s$^{-1}$}

\def\Fermi{\textit {Fermi}}
\def\Swift{{\it Swift}}
\def\Chandra{{\it Chandra}}
\def\XMM{{\it XMM}}

\def\actaa{Acta Astron.}

\def\mybf{}
\def\nbf{}
\def\obf{}

\usepackage{draftcopy}

\draftcopyName{}{24}

\begin{document}

\submitted{}
\accepted{August 13, 2016}

\title{A Luminous Gamma-ray Binary in the
Large Magellanic Cloud}

\author{R.~H.~D. Corbet\altaffilmark{1}, L. Chomiuk\altaffilmark{\obf{2}}, M.~J. Coe\altaffilmark{{\obf 3}}, 
J.~B. Coley\altaffilmark{{\obf 4}}, G. Dubus\altaffilmark{{\obf 5}}, \\
P.~G. Edwards\altaffilmark{{\obf 6}}, P. Martin\altaffilmark{{\obf 7}}, V.~A. McBride\altaffilmark{{\obf 8,9}}, 
J. Stevens\altaffilmark{{\obf 6}}, J. Strader\altaffilmark{{\obf 2}}, L.~J. Townsend\altaffilmark{{\obf 8}}, \\
A. Udalski\altaffilmark{{\obf 10}}
}

\altaffiltext{1} {University of Maryland, Baltimore County, and
X-ray Astrophysics Laboratory, Code 662 NASA Goddard Space Flight Center, Greenbelt Rd., MD 20771, USA.
Maryland Institute College of Art, 1300 W Mt Royal Ave, Baltimore, MD 21217, USA.}

\altaffiltext{{\obf 2}} {Department of Physics and Astronomy, Michigan State University, East Lansing, MI 48824, USA.}

\altaffiltext{{\obf 3}} {University of Southampton, School of Physics and Astronomy, Southampton SO17 1BJ, UK.}

\altaffiltext{{\obf 4}} {NASA Postdoctoral Program, and Astroparticle Physics Laboratory, 
Code 661 NASA Goddard Space Flight Center, Greenbelt Rd., MD 20771, USA.}

\altaffiltext{{\obf 5}} {Institut de Plan\'{e}tologie et d'Astrophysique de Grenoble, Univ. Grenoble Alpes, CNRS, F-38000 Grenoble, France.}

\altaffiltext{{\obf 6}} {Commonwealth Scientific and Industrial Research Organisation Astronomy and Space Science, PO Box 76, Epping, New South Wales 1710, Australia.}

\altaffiltext{\obf{7}} {Institut de Recherche en Astrophysique et Plan\'etologie, Universit\'e de Toulouse, CNRS, F-31028 Toulouse cedex 4, France.}

\altaffiltext{{\obf 8}} {{\obf Department of Astronomy, University of Cape Town, Private Bag X3, Rondebosch, 7701, South Africa.}}

\altaffiltext{{\obf 9}} {South African Astronomical Observatory, PO Box 9, Observatory, 7935, South Africa.}
 
\altaffiltext{{\obf 10}} {Warsaw University Observatory, Al. Ujazdowskie 4, 00-478 Warszawa, Poland.}

\begin{abstract}
Gamma-ray binaries consist of a neutron star or a black hole
interacting with a normal star to produce gamma-ray emission that
dominates the radiative output of the system. Only a handful of such
systems have been previously discovered, all within our Galaxy.
Here we report the discovery with the \Fermi\ Large Area Telescope (LAT)
of a luminous gamma-ray binary in the Large Magellanic Cloud {\mybf from a search
for periodic modulation in all sources in the third \Fermi\ LAT catalog.}
{\mybf This is the}
first such system to be found outside the Milky Way. The system has an
orbital period of 10.3 days and is associated with a massive O5III
star located in the supernova remnant \dem, previously identified as the
candidate high-mass X-ray binary (HMXB) \cxou.  X-ray and radio emission are
also modulated on the 10.3 day period, but are in anti-phase with the
gamma-ray modulation.  Optical radial velocity measurements suggest
that the system contains a neutron star.  The source is significantly
more luminous than similar sources in the Milky Way at radio, optical,
X-ray and gamma-ray wavelengths.  The detection of this extra-galactic
system{\mybf, but no new Galactic systems} raises the possibility that the predicted number of gamma-ray
binaries in our Galaxy have been overestimated, and that HMXBs may be
born containing relatively slowly rotating neutron stars.
\end{abstract}

\section{Introduction}
Although hundreds of interacting binary systems are known
X-ray emitters \citep{Liu2006,Liu2007}, very few systems produce
detectable gamma-ray emission. 
{\mybf We here classify gamma-ray binaries as those systems
where most of the electromagnetic output of the system
is at gamma-ray energies.} 
In order to generate gamma-rays,
non-thermal emission mechanisms are required such as
the particle acceleration in a shock between
the wind from a rapidly rotating neutron star and its companion \citep{Dubus2006},
where the Fermi mechanism \citep{Marcowith2016} may accelerate particles to high energies, 
or in the high-velocity jets from an accreting black-hole ``microquasar'' {\mybf \citep[e.g.][]{Mirabel1998}.
Gamma-ray emission has been detected from the microquasar
Cygnus X-3 \citep[e.g.][]{FermiLAT2009,Tavani2009,Corbel2012}
and possibly from Cygnus X-1 \citep{Bodaghee2013}. However these two
sources can be viewed as ``gamma-ray emitting X-ray binaries" 
since emission predominantly occurs at X-ray energies \citep{Dubus2015}.
}

Some evolutionary models of HMXBs predict that 
these systems pass through such a brief gamma-ray binary phase shortly after
the formation of a short {\mybf rotation} period neutron star in a supernova explosion \citep{Meurs1989}
in a binary system with an O or B spectral type companion.
A population of gamma-ray binaries is thus expected, the observable number
depending on factors including the duration of the gamma-ray binary
phase and the gamma-ray luminosity. 
{\mybf
From their binary population synthesis study
\citet{Meurs1989} predicted about 30 binaries containing neutron stars during
their pulsar phase which could thus be gamma-ray binaries.
Following the launch of the {\it Fermi Gamma-ray Space Telescope} mission in 2008, its
Large Area Telescope \citep[LAT;][]{Atwood2009} was used to confirm
GeV gamma-ray emission from the systems PSR B1259-63, \lsi, and \ls5039\
which had been suspected to be members of this class of object
\citep[e.g.][and references therein]{Dubus2013}.
All three of these systems display modulation of their gamma-ray fluxes on their orbital periods.
This suggests that detection of periodic emission from a gamma-ray source can be a powerful
way to find new binaries.
From an earlier search for periodic modulation in cataloged \Fermi\ LAT sources 
a 16.5 day period was found from \J1018\ which multi-wavelength observations
confirmed to be a gamma-ray binary \citep{Corbet2011,FermiLAT2012}.
These initial results suggested that the predicted population of \sqig30 Galactic gamma-ray
binaries might indeed exist. However, since the discovery of \J1018\ in 2011
no additional gamma-ray binaries had been found with \Fermi, possibly calling into question the
number of such sources in the Galaxy.
On the other hand, we note that the gamma-ray binary HESS J0632+057, although detectable at higher TeV
energies, is at most only a faint source at the GeV energies accessible to {\nbf\Fermi} \citep{Malyshev2016},
and the pulsar PSR J2032+4127 might become a gamma-ray binary at the periastron passage of its
highly eccentric 20-30 year long orbit around its Be star companion \citep{Lyne2015}.
}

We describe here our program to search for additional gamma-ray binaries from the detection
of periodic modulation in the \Fermi\ LAT light curves of sources in the {\mybf third} \Fermi\ source catalog. 
This {\mybf has} resulted in the discovery of periodic modulation
with a 10.3 day period from the direction of the Large Magellanic Cloud (LMC).
{\mybf The LMC is a neighbor galaxy of the Milky Way located \sqig50 kpc from the
Earth \citep{Macri2006,Piet2013,deGrijs2014}}. 
This source was then localized to
a position that suggested identification with the ``P3'' point-like source in a LAT survey of
the LMC by \citet{Ackermann2016}. We next identified the counterpart as a previously proposed
high-mass X-ray binary (HMXB) \cxou\ \citep{Seward2012}.
A previous observation of the {\mybf supernova remnant (SNR)} \dem\ surrounding \cxou\ with ATCA \citep{Bozzetto2012}
had revealed a point-like source which had been interpreted, following an
earlier suggestion \citep{Bamba2006} based on {\it XMM-Newton} data, as a pulsar-wind nebula.

To confirm the identification of \cxou\ as the counterpart of the periodic gamma-ray
source
we obtained X-ray and radio observations of \cxou, using the \Swift\ satellite and
the Australia
Compact Telescope Array (ATCA) respectively,
and found that the
X-ray and radio fluxes from this source are
also modulated on the 10.3 day period. \cxou\ has a previously identified O5III(f) counterpart
\citep{Crampton1985, Seward2012}.
This V = 13.5 optical counterpart had previously been
reported to show up to 30 km\,s$^{-1}$ velocity differences from day to day \citep{Seward2012,Crampton1985}.
We therefore obtained optical radial velocity measurements with the Southern Astrophysical Research (SOAR) 4.1 m
and South African Astronomical Observatory (SAAO) 1.9 m telescopes
which allowed us to determine that the source probably contains
a neutron star. {\mybf This is thus the first
gamma-ray binary to be found outside the Milky Way
and also the first gamma-ray binary to be found with the \Fermi\ LAT since \J1018.}

We describe LAT observations and our program to search for modulated
gamma-ray sources in Section \ref{sect:gray_obs}. The \Swift\ X-ray observations
of \cxou\ are presented in Section \ref{sect:xray_obs} and the ATCA observations
in Section \ref{sect:radio_obs}. Optical spectroscopy to obtain radial velocity
measurements is described in Sections \ref{sect:soar_obs} and \ref{sect:saao_obs}
and optical photometry obtained
with Optical Gravitational Lensing Experiment
(OGLE) is described in Section \ref{sect:ogle_obs}. 
The discovery of periodic gamma-ray emission from the direction of the LMC is presented in
Section \ref{sect:gray_results}, the identification of the counterpart as \cxou\ in \dem\ from the
detection of modulated X-ray and radio emission is
given in Sections \ref{sect:xray_results} and \ref{sect:radio_results} respectively. Constraints on
the system from optical radial velocity measurements are given in Section \ref{sect:rv_results} and
OGLE upper limits on photometric modulation in Section \ref{sect:ogle_results}.
The nature of LMC P3/\cxou\ and
the implications of the discovery of a gamma-ray binary outside the Milky
Way, but the lack of new sources in our Galaxy, for the overall population of gamma-ray binaries
are discussed in Section \ref{sect:discuss} with an overall conclusion in Section \ref{sect:conclude}.
{\mybf Unless otherwise stated, uncertainties are given at the 1$\sigma$ level.
For luminosity calculations we adopt a distance of 50 kpc.
}

\section{Observations and Analysis}

\subsection{Gamma-ray Observations and Analysis\label{sect:gray_obs}}

All gamma-ray observations were obtained with the LAT on board the \Fermi\ satellite \citep{Atwood2009}.
The LAT is a pair conversion telescope sensitive to gamma-ray photons with energies
between \sqig20 MeV to $>$ 300 GeV.
The LAT data used here were obtained between {\mybf 2008 August 5} and {\mybf 2016 March 24} (MJD 54,683
to 57,471). 
{\mybf The initial search for gamma-ray binaries was performed with a somewhat shorter
data set obtained with the same start date up to 2015 August 27 (MJD 57,261). }
Analysis was performed using version v10r0p5 of the {\nbf\Fermi} Science
Tools, with Pass 8 ``Source'' class, front plus back data, for an energy range
of 100 MeV to 300 GeV.

The third \Fermi\ LAT catalog \citep[``3FGL'',][]{Acero2015} contains 3033 sources.
To search for gamma-ray binaries we created light curves of all 3FGL sources and calculated
power spectra of these to investigate the presence of periodic modulation.
{\mybf
For the strongest peak in each power spectrum
the False Alarm Probability \citep[FAP,][]{Scargle1982}, the estimated probability 
of a signal reaching a power level by chance under the assumption of white noise,
was calculated. This FAP takes into account the number of independent frequencies
searched, but does not include the effect of searching for periodicity in multiple
sources.}
The light curves{\nbf, covering an energy range of 100 MeV to 500 GeV,} were
created using a modified version of aperture photometry
where the probability that a photon comes from
a source of interest is summed, rather than simply the
number of photons {\mybf \citep[][and references therein]{FermiLAT2012}}. To estimate the probabilities,
models were created for each source using the 3FGL catalog
using sources within a 10 degree radius and \texttt{make3FGLxml}. Photon
probabilities were calculated using \texttt{gtsrcprob} and
then summed for a 3 degree radius aperture centered on each source.
{\mybf Time bins of 500s were used for all sources.}

Power spectra of weighted-photon aperture photometry LAT light curves were calculated weighting each
data point's contribution by its relative exposure, after first subtracting
the mean count rate. This is required because of the large exposure changes
from time bin to time bin which are particularly apparent because of the short
time bins {\mybf \citep{FermiLAT2009}}. {\mybf For each source the calculated power spectrum covered a period range from 0.05 days
(1.2 hrs) to the length of the light curve, i.e. \sqig2788 days for the full data set and
\sqig2578 days for the initial search.
The power spectra were oversampled by a factor of 5 compared to the nominal
frequency resolutions of \sqig1/2788 days$^{-1}$ and \sqig1/2578 days$^{-1}$ respectively.
} 
Since background is not fitted for each bin, artifact signals can be
seen at \Fermi's orbital period, the survey period at twice this,
one day, {\mybf the Moon's 27.3 day sidereal period},
the 53 day precession period of the \Fermi\ satellite, {\mybf and one quarter of a year} 
\footnote{http://fermi.gsfc.nasa.gov/ssc/data/analysis/LAT\_caveats\_temporal.html}.

After detecting likely modulation from the region of the LMC,
light curves were then generated by using a model for the LMC \citep{Ackermann2016}.
Weighting each photon by its probability of coming from a
source of interest has been found to increase the signal-to-noise
level of the light curve \citep{FermiLAT2012,Kerr2011}. However, it can reduce the apparent
modulation of the light curve compared to its actual variability.
When a source is brighter than its average level the probability that
a photon came from the source is underestimated, conversely the probability
will be overestimated when the source is fainter than its mean
level. This effect is especially pronounced for a faint source in the
presence of brighter emission from neighboring sources or overall
background level. For this reason likelihood analysis provides a 
much more reliable estimate of the amplitude of source variability.

\subsection{X-ray Observations and Analysis\label{sect:xray_obs}}

The \textsl{Swift} X-ray Telescope \citep[XRT;][]{Burrows2005} is a Wolter I X-ray imaging telescope 
with a focal length of 3.5\,m fitted with a CCD chip covering a region of 23.6$\arcmin$$\times$23.6$\arcmin$. 
The energy resolution FWHM in the XRT at the time of launch was $\sim$140\,eV at 5.9\,keV.  Sensitive to X-rays 
ranging from 0.3 to 10 keV, the effective area of the XRT is $\sim$125\,cm$^{2}$ at 1.5\,keV.  
At 8.1\,keV, the effective area of the XRT is $\sim$20\,cm$^{2}$.  The count rate for a 
source of 1\,mCrab is 0.7\,counts s$^{-1}$ \citep{Hill2004}.

XRT observations of \cxou\ took place from 2015 {\mybf November} 21 to 2016 {\mybf January} 19 
(MJD\,57,347--57,406) with exposures ranging from $\sim$1.1\,ks to $\sim$4.9\,ks.  Table~\ref{table:xrt} 
gives the observation log. 
The data were reduced and analyzed using the \textsl{Swift} XRT product
generator \citep{Evans2007} and the standard criteria given in the \textsl{Swift} XRT Data Reduction Guide \citep{XRTGuide}.  
These procedures are described below.

\cxou\ was observed in Photon Counting \citep[PC;][]{Hill2004} mode with a readout time of 2.5\,s adopting the standard grade filtering (0--12 for PC).  Data were reduced and screened using \texttt{xrtgrblc} and \texttt{xrtgrblcspec} in HEAsoft v.6.19.  The data were reprocessed with the XRTDAS data pipeline package \texttt{xrtpipeline} using the standard filtering procedure to apply the newest calibration and default screening criteria.  The source spectra were extracted from count dependent regions generated by \texttt{xrtgrblc}.  
An unrelated field source was found at 
(J2000) R.A. = 05$^{\rm h}$35$^{\rm m}$47$^{\rm s}_.$4, Dec = $-$67$\degr$31$\arcmin$55$\arcsec_.$4, 
which was excised from the background extraction. 
Because of the presence of the SNR we restrict our X-ray analysis to energies
above 2 keV where SNR emission is reduced \citep{Bamba2006}.

The ancillary response files, accounting for vignetting, point spread function correction and different extraction regions, were generated and corrected for exposure using the FTOOLS packages \texttt{xrtmkarf} and \texttt{xrtexpomap}, respectively.  
{\mybf We find the count rates in the observations to be $\sim$(1.1--3.4) $\times$ 10$^{-2}$\,counts s$^{-1}$
(full energy range, source plus background), which is significantly less than the count rate of
0.5\,counts s$^{-1}$ where pileup becomes
important.}

Individual spectra were not useful for analysis, as each spectrum was found to have 22--123 counts in the 0.3--10\,keV energy band.  A cumulative spectrum was therefore extracted, which has a total of 784 counts.  Photons in the 2--10\,keV energy band were considered to reduce contamination
from the DEM L241 supernova remnant \citep{Bamba2006}.  This gave a total of 119 counts for this energy band 
and the total exposure is $\sim$34.9\,ks.  Data in the spectral files produced by \texttt{xselect} were
further processed using \texttt{grppha}
which is designed to define the binning, quality flags and systematic errors of the spectra and used the quality flag to further eliminate bad data from the PHA files.  Initially, we grouped the bins to ensure a minimum of 20 counts to fit the spectra using $\chi^2$ statistics.  However, insufficient bins were produced in the resulting spectrum.  
Due to the small number of counts, we therefore used the ``C'' statistic {\mybf \citep{Cash1979}} for the spectral analysis.  
The cumulative spectrum was grouped to have 5 counts per bin.
The spectra were analyzed using \texttt{XSPEC v12.9.0k}.  We made use of the \texttt{XSPEC} convolution model \texttt{cflux} to calculate the fluxes and associated errors of \cxou.

\subsection{Radio Observations and Analysis\label{sect:radio_obs}}

Radio observations were obtained using the Australia Telescope
Compact Array \citep[ATCA;][]{Wilson2011}.
Dedicated follow-up observations were made between 2015 {\mybf November} 29 and 2016 {\mybf March} 12 (MJD 57,355 to 57,459, see
Table \ref{table:atca})
with observations centered at 5.5 and 9.0 GHz, with 2 GHz bandwidths for both bands.
The ATCA, which consists of six 22m-diameter antennas, was in several different array configurations over this period:
1.5A (minimum baseline 153 m, maximum baseline 4.5 km),
750C (46 m, 5.0 km),
EW352 (31 m, 4.4 km),
and 6B (214 m, 6.0 km).
Observations were reduced following standard procedures in Miriad \citep{Sault1995}, with the flux density scale set by
observations of calibrators PKS 1934-638 and/or PKS 0823-500.
Initially, mosaiced observations were made covering both the nominal position of the gamma-ray source P3, and
the proposed X-ray counterpart \cxou.
Observations differed in length, hour-angle coverage, angular resolution, and sensitivity to extended radio emission
in the vicinity, resulting in a heterogeneous data set. 
The final observations of the series were conducted as targeted observations rather than mosaics
of the region, with both a source inside the LAT error region and \cxou\ radio sources observed together with the phase calibrator
PKS 0530-727.

We also examined the Australia Telescope On-line Archive (atoa.atnf.csiro.au) for other observations of this region.
Mosaiced observations were made in 2002 at 4.8 GHz with a 128 MHz bandwidth
with simultaneous, but spatially undersampled, observations at 8.4 GHz.
Investigation of this data did not reveal any counterpart stronger
than several mJy, with the image dominated by the
nearby bright H\,{\sc ii} region PKS 0535$-$676, $\sim$1.5\,Jy at 4.8\,GHz.

\subsection{Optical Observations and Analysis\label{sect:optical_obs}}
\subsubsection{SOAR\label{sect:soar_obs}}

Long-slit spectra of the optical counterpart of \cxou\ were acquired using the 
Goodman spectrograph \citep{Clemens2004} on the 4.1 m SOAR telescope. 
All observations used a 1.03\arcsec\ slit and a 2100 l mm$^{-1}$ grating, yielding a resolution of 0.9\AA\ over a wavelength range $\sim$ 4300-4970 \AA. At each epoch 20 minutes of exposure time was obtained. The spectra were optimally extracted and calibrated in the usual manner.

Qualitatively the spectra appear identical to those presented of the system in previous work \citep{Crampton1985,Seward2012}, with absorption lines of H and He and the Bowen blend in emission. We determined barycentric radial velocities through cross-correlation with a field O star taken with a similar setup around the region of the He\,II line at 4542 \AA.

For this paper, the first observations were obtained on 2015 {\mybf December} 10 (UT) and the last on 2016 {\mybf April} 16 
(Table \ref{table:soar}) 
comprising 13 separate epochs listed as Barycentric Julian Dates on the TDB system \citep{Eastman2010}. 
Nonetheless, due to the long period of the system the phase coverage is still not optimal.

\subsubsection{SAAO Observations\label{sect:saao_obs}} 
Spectra were obtained using the newly upgraded spectrograph at the Cassegrain focus of the 1.9 m Radcliffe telescope at SAAO. The spectrograph's collimator, optics and detector were all replaced during the second half of 2015. All associated software was also rewritten during the upgrade.

Blue spectra were obtained using two gratings: The first has a broad spectral range of 800 \AA\ and a resolution of 1 \AA. The second has a {\mybf second} order spectrum in the blue, which allowed for an improved resolution of 0.5 \AA\ while sacrificing some spectral range and throughput. Observations using the two gratings were performed as part of science verification tests.

Three observations were obtained during November and December 2015 after the initial testing of the upgraded spectrograph.
Exposure times of 1200 s were used for both gratings, achieving a good signal-to-noise ratio in each spectrum.

\subsubsection{OGLE Photometry\label{sect:ogle_obs}}

Optical I-band data from OGLE \citep{Udalski2015}
Phase IV  were used to investigate the long-term and periodic behavior
of the optical counterpart in this system.
The source is identified as LMC518.09.16278 within the OGLE
IV phase of the project. The data cover a period of several years starting at MJD 55,260 and continuing 
to MJD 57,339. The coverage was almost nightly for the first 2-3 years, but more recently the frequency of observations has been reduced to approximately once every 3-4 nights.

\section{Results\label{sect:results}}

\subsection{Gamma-ray Results\label{sect:gray_results}}

{\mybf
We examined the power spectra of the light curves of all 3FGL sources to search for evidence
of periodic modulation that could be the sign of a binary system.
The binary systems \lsi, \ls5039, and \J1018\ were all detected at extremely
high levels of statistical significance (ratio of peak power to mean power $>$ 100; FAP $<$ 10$^{-9}$).
In addition, these binary signals were also often strongly seen in the power 
spectra of nearby sources.
For candidate new binaries our threshold for further investigation was for a source
to have a peak power $\ge$ 18 $\times$ mean power level (FAP $<$ 5$\times$10$^{-4}$)
and for the period not to coincide with a known artifact. Some sources were found with modulation above
this threshold but which could be interpreted as due to non-periodic modulation from an
active galactic nucleus (AGN). Such sources could be identified if the source was already associated with
an AGN, if the source was located in a region where the presence of an early type star would be
unlikely such as far from the Galactic plane, or the peak appeared to be part of underlying low frequency noise and so due to
non-periodic variability.
A five sigma detection of a binary nominally arises with a relative peak height of $>$25.
However, to ensure that a signal does not have some other cause such
as an unknown artifact, full confidence
in the discovery of a new binary requires the identification of a counterpart
at other wavelengths and the detection of the same period in the counterpart.
}

From the power spectra of the 3FGL light curves we noted a peak near 10.3 days 
(height \sqig22$\times$ mean power with an FAP of \sqig$10^{-5}$) 
from 3FGL J0526.6-6825e. This is the LMC, 
that is treated as a single object in the 3FGL catalog.
A ``difference" image,
made by subtracting an image of 10.3 day minimum phase only from an image of 10.3 day maximum
phase only, showed the modulation to be localized, and somewhat offset
from the catalog position at roughly right ascension = 84.0\deg, declination = -67.55\deg\ (J2000).
The modulation center was near ``P3'', an unassociated source recently found in LAT observations of the
LMC by \citet{Ackermann2016}, who noted several potential counterparts, including the SNR DEM L241 just outside the localization error region.

\begin{figure}
\includegraphics[width=7.25cm,angle=270]{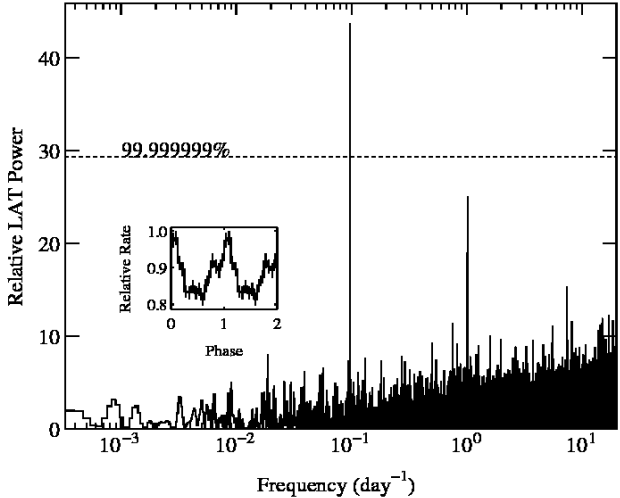}
\caption{Power spectrum of the weighted-photon LAT light curve (E $>$ 100 MeV)
of {\nbf\Fermi} LMC P3. The strongest peak is at the proposed 10.3 day
orbital period of the system. The second highest peak is a common
artifact in LAT light curves near 1 day. The horizontal line shows
the indicated false alarm probability level.
The inset shows the light curve folded on the 10.3 day period.
For clarity two cycles are shown.}
\label{fig:p3_power}
\end{figure}

\begin{figure}
\includegraphics[width=8.75cm]{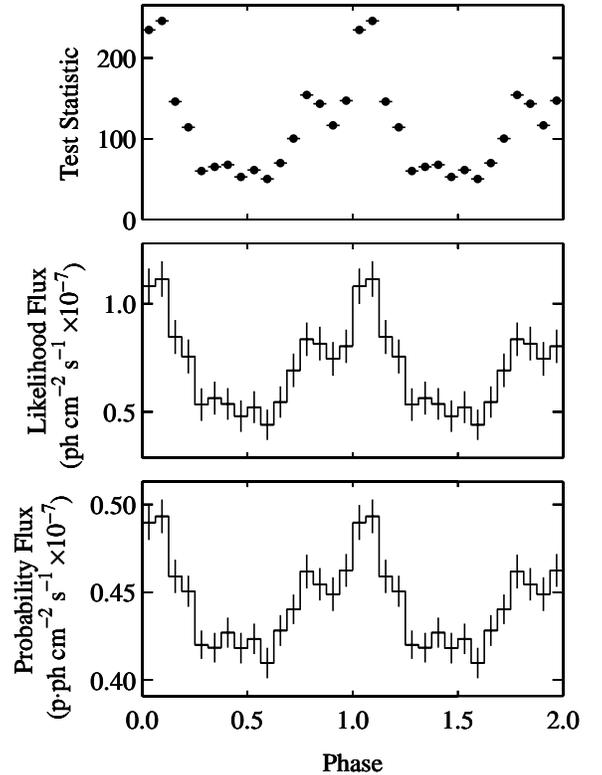}
\caption{Phase-resolved LAT observations {\nbf(E $>$ 100 MeV)} of LMC P3.
From top to bottom:
(i) Test Statistic (TS) \citep{Mattox1996} from likelihood analysis,
(ii) flux from phase-resolved likelihood analysis, 
(iii) folded probability-weighted aperture photometry.
}
\label{fig:lat_fold}
\end{figure}

\begin{figure}
\includegraphics[width=8.75cm]{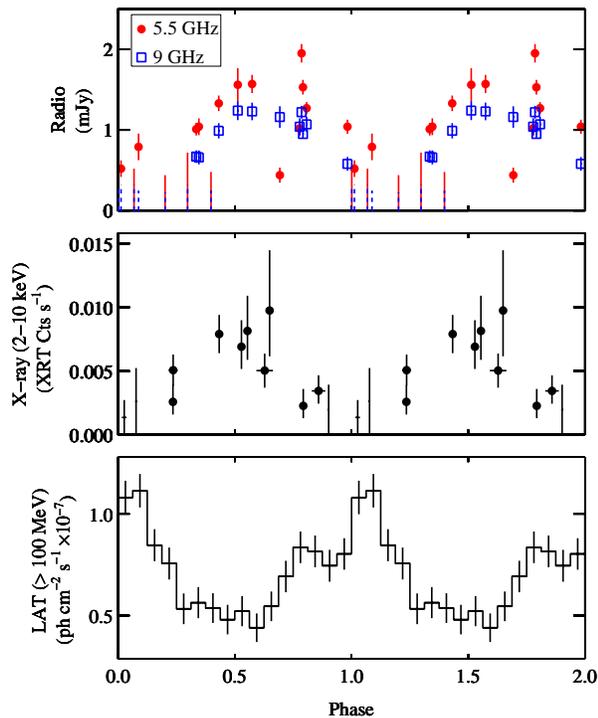}
\caption{Radio (top), X-ray (middle), and gamma-ray (bottom) fluxes
from LMC P3/\cxou\ folded on the 10.3 day period. Phase zero is the time of
maximum flux for sinusoidal modulation of the gamma-ray flux and corresponds
to MJD 57,410.25. 
For the radio
flux densities the lines extending down to zero indicate 
4 $\sigma$\ upper limits. For the X-ray the three lines extending down to zero show 3 $\sigma$\ upper limits.
}
\label{fig:p3_fold}
\end{figure}

The spectral-spatial model for gamma-ray emission from the LMC \citep{Ackermann2016} was used
to create probability-weighted aperture photometry light curves, including a position
centered on DEM L241.  The 10.3 day peak in the power
spectrum increased to \sqig44$\times$ mean power (Figure \ref{fig:p3_power})
with an {\mybf FAP}  
of less than 10$^{-10}$ strongly suggesting this was the location of the modulated gamma-ray source.
The period was found to be 10.301 $\pm$ 0.002 days with epoch of maximum flux for sinusoidal
modulation 
of MJD 57,410.25 $\pm$ 0.34 (Figure \ref{fig:lat_fold}).
We also performed a likelihood fit to phase-resolved LAT spectra which 
revealed a large modulation amplitude with a profile that is more complex than
purely sinusoidal (Figure \ref{fig:p3_fold}). 
For the phase-resolved likelihood analysis we again used the LMC model \citep{Ackermann2016}
for a 10 degree radius centered on P3 for an
energy range of 100 MeV to 300 GeV. The parameters of all sources, apart from
P3 were frozen to their previously determined best fit values.
Leaving the power law index of P3 free for the phase-resolved likelihood fits
resulted in unreasonably steep spectra with power-law indices of $\ge$ 5 for phases of lower flux.
We therefore also froze the power-law index to the value of 2.77 from \citet{Ackermann2016}.
To obtain a more robust measurement of spectral variability it may be necessary
to rederive a model for this complex region using Pass 8 data and the knowledge of the precise
location of P3. Such an analysis is beyond the scope of this paper.

\subsection{X-ray Results\label{sect:xray_results}}

The XRT observations of \cxou\ monitor the source for more than three orbital periods.
We binned the light curves to a resolution of one bin per observation to investigate the X-ray orbital modulation of the system.  
The \Swift\ observations revealed strong, approximately sinusoidal, modulation on the 10.3 day
gamma-ray period (Figure \ref{fig:p3_fold}).
However, X-ray minimum occurs near the phase of gamma-ray maximum.

\begin{figure}
\includegraphics[width=5.65cm,angle=270]{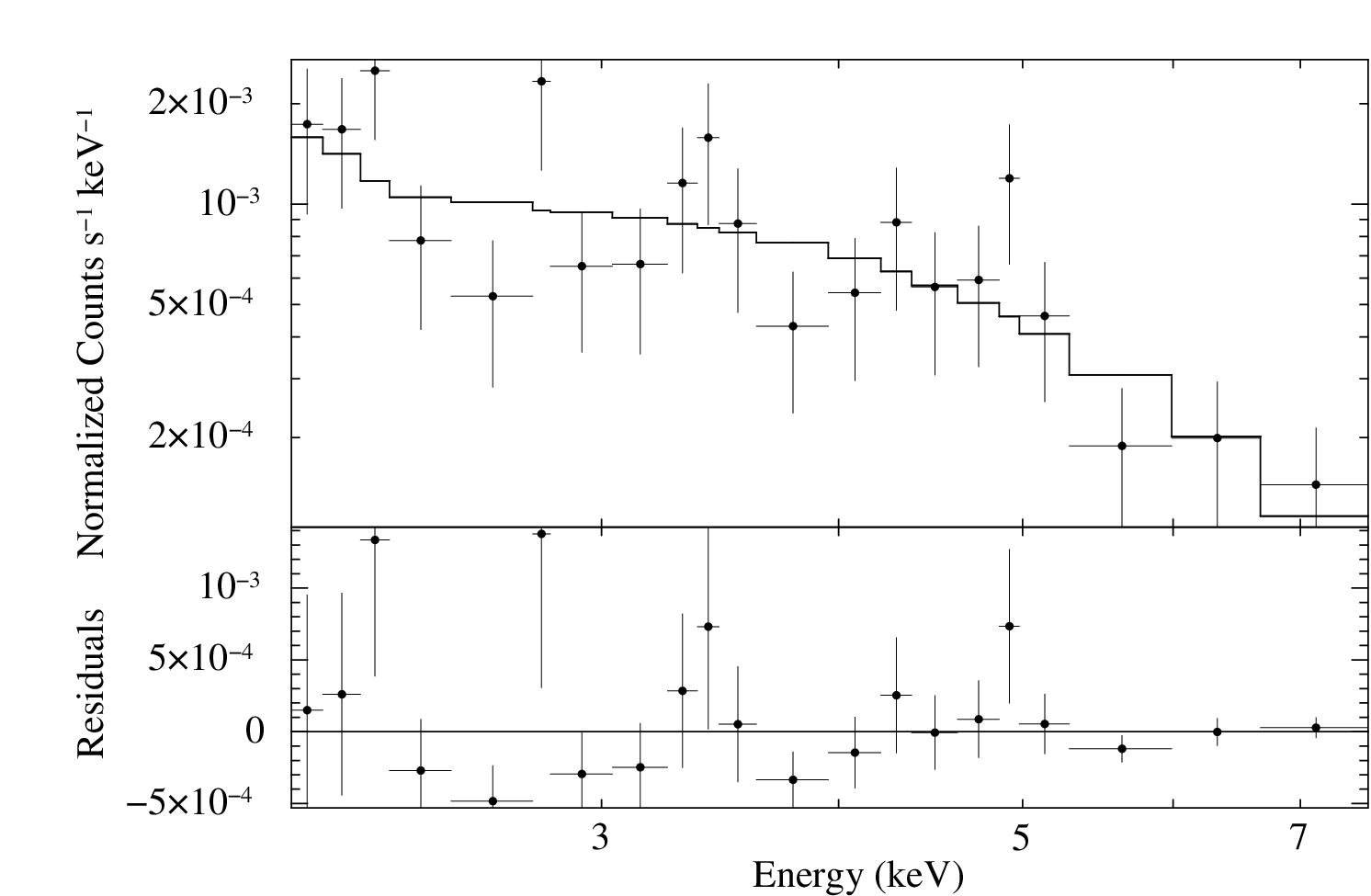}
\caption{Cumulative Swift XRT spectrum of \cxou}
\label{fig:xrt_spectrum}
\end{figure}

To fit the cumulative {XRT} spectrum, we used several models that are used to describe systems that host a neutron star: 
a power law (see Figure \ref{fig:xrt_spectrum}), a power law with a high-energy cutoff (\texttt{highecut$\times$power} in XSPEC), and a cutoff powerlaw (\texttt{cutoffpl} in XSPEC).  All models were modified by an absorber that fully covers the source using appropriate 
cross sections \citep{Balucinska1992} and abundances \citep{Wilms2000}.

The model that provides a good fit {\mybf (C statistic of 20.15 for 19 degrees of freedom)} to the data is a power law with photon index $\Gamma$=1.3 $\pm$ 0.3 modified by a fully covered absorber (\texttt{tbabs} in XSPEC).
{\mybf 
We find the unabsorbed flux of \cxou\ to be (3.2$\pm$0.4)$\times$10$^{-13}$\,erg\,cm$^{-2}$\,s$^{-1}$ in the 2.0--7.5\,keV band.
Our photon index
is thus consistent with the values of 1.51 - 1.62 found by \citet{Bamba2006}
and 1.28$\pm$0.08 reported by \citet{Seward2012}.}  

We find that the neutral hydrogen column density for the fully covered absorption could not be accurately constrained.  
This is not surprising as only energies above 2\,keV are included in the analysis.  
Therefore, we froze $N_{\rm H}$ to 1.9$\times$10$^{21}$\,atoms cm$^{-2}$, which was found from \textsl{Chandra} data \citep{Seward2012}.
This value of the fully covered absorber is comparable with the Galactic HI value
given in the Leiden/Argentine/Bonn survey \citep{Leiden2005}, which is 1.62$\times$10$^{21}$\,atoms cm$^{-2}$.  
Therefore, we assume that the fully covered absorber is interstellar in origin.  
While a good fit does not require a high-energy cutoff, which is typically found in accreting pulsars,
{\mybf our spectra are limited to energies below 10 keV and such cutoffs are often seen at higher energies}
\citep{Coburn2002}.

{\mybf
\subsection{Comparison with Previous X-ray Observations}
In Table \ref{table:all_xray} we summarize our current and also previous X-ray observations
of \cxou.
We note that the two \Chandra\ observations from \citet{Seward2012}
showed
an increase in the flux {\mybf between 0.5 to 5 keV} 
from (3.71 $\pm$ 0.10) to (4.70 $\pm$ 0.12) $\times$ 10$^{-13}$ \ergcm2s.
The observation times correspond to phases of \sqig0.17 and \sqig 0.26 and the flux increase
is thus consistent with the behavior we observe at these phases with the \Swift\ XRT.
However, the \XMM\ observation of \citet{Bamba2006} was obtained at a phase of \sqig0.6
which corresponds to orbital maximum but did not show a particularly high flux.
This could be indicative of cycle-to-cycle variability if confirmed by additional observations.
}

\subsection{Radio Results\label{sect:radio_results}}

A point source was found in the LAT error region of the gamma-ray source
together with the detection of a point source at the position of \cxou.
The radio flux densities folded on the 10.3 day period (Figure \ref{fig:p3_fold}) 
show modulation of the emission on the gamma-ray period, but again 
out of phase with the gamma-ray modulation. 
We note that the folded 5.5 GHz light curve is more ``noisy'' than that at 9 GHz.
In particular at phase \sqig0.68 there is a low flux 5.5 GHz point without a
corresponding decrease at 9 GHz.
The 9 GHz light curve is expected to be less susceptible to systematic effects
due to the smaller beam size at this wavelength which will result in reduced
contamination from other LMC sources detected in sidelobes of the telescope
response. {\mybf However, we note that the large scatter in 5.5 GHz fluxes
seen around the phase range \sqig0.7 -- 0.8 might also suggest cycle-to-cycle
variability.}

\subsection{Optical Results}

\subsubsection{SOAR and SAAO Optical Spectroscopic Results\label{sect:rv_results}}

\begin{figure}
\includegraphics[width=6.9cm,angle=270]{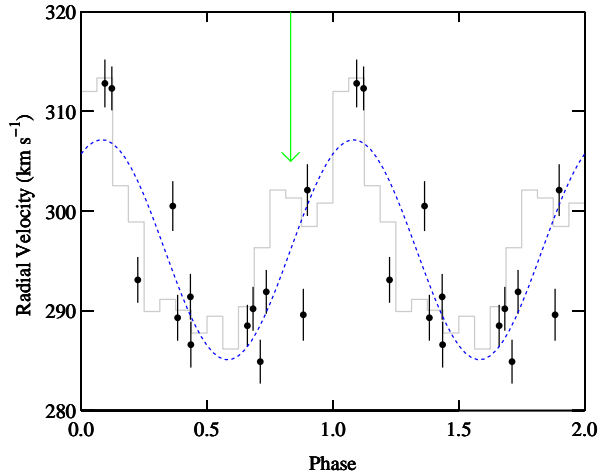}
\caption{Radial velocity measurements of the optical counterpart of \cxou\ obtained
with SOAR (plotted as filled circles with error bars) folded on the 10.3 day orbital
period. The arrow indicates the time of superior conjunction determined from a fit to the
radial velocity measurements (MJD 57408.61 $\pm$ 0.28). The gray histogram shows the gamma-ray flux measured with the LAT. 
}
\label{fig:p3_rv}
\end{figure}

The overall optical continuum and line strengths in the SOAR and SAAO spectra
showed no obvious changes from previous observations.
Radial velocities were determined from the SOAR spectra and these are shown in Figure \ref{fig:p3_rv}
folded on the 10.3 day period.
We find a clear orbital variation consistent with binary motion.

We performed circular Keplerian fits to the radial velocities with the period fixed to the value determined from analysis of the {\mybf LAT} data: $10.301$ d. Two sets of fits were performed: one with the phase fixed such that the peak gamma-ray flux occurs when the compact object is behind the O star; the other with the phase free. For the former fit, we find 
$K_2$, {\nbf the semi-amplitude of the radial velocity due to the orbital motion of the O star}, $= 5.5\pm2.7$ km s$^{-1}$, 
i.e., marginal evidence of radial velocity variations with this period and phase.

With the phase free, the best-fit parameters are: systemic velocity $295.8\pm2.0$ km s$^{-1}$ and $K_2 = 10.7\pm2.4$ km s$^{-1}$. This fit is nonetheless poor ($\chi^2/\nu$ = 70/10; rms = 5.2 km s$^{-1}$), suggesting that residual variations due to, e.g., a time-variable wind are present. The systemic velocity is consistent with that expected for the LMC disk at this location  
\citep[$277\pm20$ km s$^{-1}$;][]{vanderMarel2002}. 
We determined the uncertainties in the time of conjunction, $K_2$, and systemic velocity using a standard bootstrap. 

Constraints on the mass of the compact object in the system as a function of inclination angle ($i$) were obtained
by calculating the ``mass function'' \citep[see e.g.][]{Strader2015}, i.e. $f(M) = PK_2^3/(2\pi G) = (M_1 \sin i)^3/(M_1 + M_2)^2$, where $M_1$ and $M_2$
are the masses of the compact object and the O star respectively, and $G$ is the gravitational constant.
The mass function  is $f(M) = (1.3^{+1.1}_{-0.6}) \times 10^{-3} M_{\odot}$. 

Assuming the O star has a mass between 25 and 42 $M_{\odot}$ \citep{Seward2012,Martins2005} and a neutron star 
mass of $1.4 M_{\odot}$, the inclination is $50^{+16}_{-13}$\deg.  For a neutron star of $2.0 M_{\odot}$ the 
inclination is $35^{+11}_{-9}$\deg. We cannot rule out that the compact object is a black hole, but it would require a low 
inclination: $14^{+4}_{-3}$\deg\ for $5 M_{\odot}$ and $8 \pm 2$\deg\ for $10 M_{\odot}$.

As a check, we also fit the radial velocities with both the period and phase free. The best fit period is 10.1 d, slightly smaller than but consistent with the more precise gamma-ray period.
We also investigated eccentric orbit fits to the radial velocities. We find that for eccentricities
less than the estimated upper limit of 0.7 there is no improvement in the reduced $\chi^2$ of
the fits. Additional radial velocity measurements will be required to better
constrain or measure the eccentricity of the system. 

Very low inclinations can in principle be constrained through a comparison of the projected rotational velocity to the breakup velocity. We do not have the appropriate comparison data to make a precise estimate of the projected rotational velocity; we roughly estimate a value of about 80 km s$^{-1}$, which would suggest the inclination cannot be lower than about $6^{\circ}$, but we emphasize that this estimate is uncertain. 

In sum, the radial velocity data are consistent with a neutron star and a wide range of inclinations, with a black hole allowable if the system is very close to face on. Observations at a wider range of phases are necessary to
fully constrain the system parameters.

For the SAAO observations, although the radial velocities inferred are generally consistent with those obtained from the SOAR observations, because of small altitude-angle dependent effects on wavelength calibration we do not use the SAAO spectra here
for determination of orbital parameters. The altitude angle effects are hoped to be removed in the future.

\subsubsection{Previous SAAO Optical Spectroscopy}

Radial velocity measurements of the optical counterpart of \cxou\  were obtained by
\citet{Seward2012} in 2011 {\mybf October} and 2012 {\mybf April},
and these, along with additional measurements between 2012 Nov – 2013 Jan are presented by
\citet{Foster2012}.
These observations confirm radial velocity changes, but with significantly larger 
uncertainties of \sqig 5 to $>$ 10 \kms\ than our SOAR measurements.
Folding these measurements on the 10.3 day orbital
period gives ambiguous results due to the larger measurement uncertainties.

\subsection{OGLE Photometric Results\label{sect:ogle_results}}

\begin{figure}
\includegraphics[width=7.25cm,angle=270]{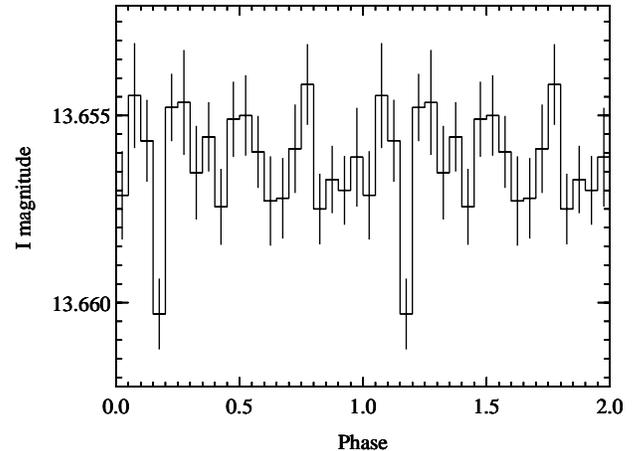}
\caption{OGLE I band light curve of the optical counterpart
of \cxou\ folded on the 10.3 day orbital period.}
\label{fig:ogle_fold}
\end{figure}

The OGLE light curve folded on the orbital period of \cxou\ is shown
in Figure \ref{fig:ogle_fold}. 
There is no statistically significant evidence for any orbital modulation in the folded OGLE data.
This does not strongly constrain the
system parameters as the expected
tidal distortion of the O star due to a 1.4\msun\ companion would be small.
We note that for the gamma-ray binary LS 5039 which has a 3.9 day period any photometric
variability is below 2 millimagnitudes \citep{Sarty2011}. Thus the lack of detection of photometric
variability for \cxou{\mybf ,} which has a longer period and a giant rather than main-sequence
primary{\mybf ,} is not surprising. 

\section{Discussion}\label{sect:discuss}

\subsection{Properties and Nature of LMC P3/\cxou\label{sect:discuss_properties}}

The modulation of the X-ray and radio fluxes of \cxou\ on the gamma-ray
period confirms that it is the counterpart of LMC P3.
The detection of radio emission is unusual for an HMXB but common for gamma-ray
binaries \citep{Dubus2015}.
The gamma-ray to X-ray luminosity ratio and identification with a massive star firmly classify P3 as a gamma-ray binary. 
The anti-correlation of the X-ray and gamma-ray orbital modulations is similar to that seen in the other two systems with O star 
companions, LS 5039 and \J1018\ \citep{Dubus2015,An2015,Hadasch2012}. 
With a 10.3 day orbit and an O5III companion, Kepler's third law requires that the orbital eccentricity is less than 0.7 for the compact object to avoid hitting the surface of its 15 $R_\odot$  companion at periastron. That these three systems show regular modulations when the three gamma-ray binaries with Be-star companions (\lsi, HESS J0632+057, PSR B1259-63) show significant orbit-to-orbit variability
\citep{Hadasch2012,Caliandro2015,Aliu2014}
is likely to be related to the lack of a circumstellar disk around the O star. 

{\mybf
The compact object in the system, which was presumably formed in the supernova explosion that also created \dem,
appears to be a neutron star. 
Thus the progenitor of the SNR, which has been proposed
to have had a mass $\gtrsim20 M_\sun$ \citep{Bamba2006} was still below the threshold for the formation
of a black hole as considered by \citet{Seward2012}.
}

The process resulting in gamma-ray emission must be steady, given the lack of evidence for long-term X-ray and gamma-ray variability. 
As with the other gamma-ray binaries, the level of emission and modulations are more likely imprinted by the interaction of a pulsar wind with the stellar wind of the O-star companion. 
The modulations are thought to be due to a combination of anisotropic inverse Compton scattering and relativistic Doppler boosting, with maximum high-energy gamma-ray emission at superior conjunction \citep{Dubus2015}. 
However, we note that gamma-ray maximum occurs somewhat after superior conjunction (Figure \ref{fig:p3_rv}), raising the possibility
that the system may have an eccentric orbit.
More precise orbital parameters will be needed to test this. 
The analogy with LS 5039 and \J1018\ also suggests that P3 may be a TeV source with a modulation correlated with the X-ray 
modulation. Scaling from the TeV/X-ray ratio of LS 5039, the maximum
flux at 1 TeV is expected to be $\sim 10^{-13}\rm\,ph\,cm^{-2}\,s^{-1}\,TeV^{-1}$,
{\mybf only slightly fainter than the currently-known TeV gamma-ray sources in the LMC \citep{HESS2015},
raising the possibility that it may be detectable with H.E.S.S.}

LMC P3 is the most luminous gamma-ray binary observed yet. 
It is at least four times more luminous in GeV gamma rays and 10 times more luminous
in radio and X-rays than LS 5039 \citep{Dubus2015,Marcote2015} and  \J1018\ \citep{FermiLAT2012}.
The luminosity of the companion, a factor 1.5 from O5V to O5III \citep{Martins2005} and the orbital separations (0.1-0.4 AU) 
are comparable in all three systems.  Hence, the higher luminosity is more likely to be due to the injected power in non-thermal 
particles rather than to higher radiation or matter densities.

The pulsar spin-down power must be  $\dot{E}\geq L_\gamma \approx 4.3\times 10^{36}$ \ergs, 
{\mybf to account for the gamma-ray luminosity}. 
For comparison, this is a factor 5 greater than the spin-down power of PSR B1259-63, which apparently converts 
nearly all this power to high-energy gamma-rays over parts of its orbit \citep{Caliandro2015}. The pulsar age is also constrained to 
$\sim 10^5$ years if it is associated with the surrounding SNR \citep{Seward2012}. Assuming no magnetic field decay and a spin-down 
power given by $\dot{E}=\mu^2 \Omega^4/c^3$ \citep{Spitkovsky2006}, with $\mu$ the pulsar magnetic moment and $\Omega$ its angular 
frequency, the magnetic field cannot be higher than $4\times 10^{11}$ G and the current spin period must be shorter than 39 ms 
{\mybf in order to be consistent with $\dot{E}\geq 4.3\times 10^{36}$ \ergs\ 10$^5$ years after the birth of the pulsar.
}

{\mybf
\subsection{Population of Gamma-ray Binaries\label{sect:discuss_population}}

The lack of discovery of additional Milky Way gamma-ray binaries since \J1018, but the detection of an LMC source, suggests that we may have detected at least the majority of persistent higher-luminosity Galactic gamma-ray binaries, even if we caution that some gamma-ray binaries may be missing from the 3FGL catalog because of their low duty cycle
\citep[e.g.][]{Caliandro2015}
or because they are GeV-faint
such as HESS J0632+057 \citep[][and references therein]{Malyshev2016}.
One possibility is that we detect as gamma-ray binaries only those systems with the fastest rotating neutron stars at birth whereas most are born as slower rotators. For example, adopting a normal distribution for birth spin periods \citep{Faucher2006} with a mean of 300 $\pm$ 150 ms these neutron stars would be
very faint in the rotation-powered stage or would go directly to the 
HMXB stage with no gamma-ray emission. In the latter case, earlier predictions of an extensive Galactic population of such objects based on the birth rate of HMXBs ($\sim 10^{-3}\rm\, year^{-1}$) and the expected lifetime of gamma-ray emission 
($\sim 10^5$ years) would have been optimistic \citep{Meurs1989}.

Although LMC P3 is considerably brighter than its Galactic counterparts, we note that even this source 
would not be detectable at the \sqig780 kpc distance of the Andromeda galaxy. Thus, unless significantly
brighter sources exist, we are currently limited to gamma-ray detection of binaries within the Milky Way and its satellites.
}

\section{Conclusion}\label{sect:conclude}
{\mybf
We have discovered a luminous gamma-ray binary with a 10.3 day period located in an SNR in the LMC using the \Fermi\ LAT.
The radio and X-ray counterparts also exhibit flux modulation on this period. 
The source properties, including radial velocity measurements of the O5 III (f) counterpart,
suggest that the system contains a rapidly rotating neutron star. The system
may eventually evolve into an X-ray binary.
Further multi-wavelength observations have the potential to enable a better modeling of the system.
Deep radio and X-ray observations should be made to search for the spin period of the neutron star.
The discovery of this source suggests that the number of Galactic gamma-ray binaries may have been
overestimated, however searches for additional systems should still continue.
}

\acknowledgements
{\obf We thank C.~C. Cheung for support and important comments
during the production of this paper.}
{\mybf We also thank an anonymous referee for useful comments.}
Based on observations obtained at the Southern Astrophysical Research (SOAR) 
telescope, which is a joint project of the Minist\'{e}rio da Ci\^{e}ncia, 
Tecnologia, e Inova\c{c}\~{a}o (MCTI) da Rep\'{u}blica Federativa do Brasil, the U.S. 
National Optical Astronomy Observatory (NOAO), the University of North Carolina at Chapel Hill (UNC), 
and Michigan State University (MSU). This work was partially supported by NASA {\nbf\Fermi} grant 
NNX15AU83G. The OGLE project has received funding 
from the National Science Centre, Poland, grant MAESTRO 2014/14/A/ST9/00121 to AU. 
The Australia Telescope Compact Array is part of the Australia Telescope National Facility which is 
funded by the Australian Government for operation as a National Facility managed 
by CSIRO. JBC was supported by an appointment to the NASA Postdoctoral Program at the
Goddard Space Flight Center administered by Universities Space Research Association through a contract with NASA.
J. Strader acknowledges support from the Packard Foundation.
We thank the Swift team for undertaking observations.
{\obf The \textit{Fermi} LAT Collaboration acknowledges generous ongoing support
from a number of agencies and institutes that have supported both the
development and the operation of the LAT as well as scientific data analysis.
These include the National Aeronautics and Space Administration and the
Department of Energy in the United States, the Commissariat \`a l'Energie Atomique
and the Centre National de la Recherche Scientifique / Institut National de Physique
Nucl\'eaire et de Physique des Particules in France, the Agenzia Spaziale Italiana
and the Istituto Nazionale di Fisica Nucleare in Italy, the Ministry of Education,
Culture, Sports, Science and Technology (MEXT), High Energy Accelerator Research
Organization (KEK) and Japan Aerospace Exploration Agency (JAXA) in Japan, and
the K.~A.~Wallenberg Foundation, the Swedish Research Council and the
Swedish National Space Board in Sweden. Additional support for science analysis during the
operations phase is gratefully acknowledged from the Istituto Nazionale di Astrofisica in
Italy and the Centre National d'\'Etudes Spatiales in France.}

\input{lmcp3_references.tex}
\input{preprint_lmcp3_tables2.tex}

\end{document}

%% file: preprint_lmcp3_tables2.tex

\begin{deluxetable}{ccccccc}

\tablewidth{6.5in}

\tablecolumns{7}
\tabletypesize{\small}
\tablewidth{0pc}
\tablecaption{Swift XRT Observation Log of CXOU J053600.0-673507}
\tablehead{
\colhead{ObsID} & \colhead{Start Time (UT)} & \colhead{End Time (UT)} & \colhead{Phase$^a$} & \colhead{Exposure$^b$} & \colhead{Count Rate$^c$} & \colhead{Flux$^d$}}
\startdata
00034169001 & 2015-11-21 09:22:58 & 2015-11-21 11:57:36 & 0.898--0.908 & 2.5 & {\mybf $<$0.39$^e$} & {\mybf $<$5.2$^e$} \\
00034169002 & 2015-11-24 18:51:58 & 2015-11-24 23:02:51 & 0.227--0.244 & 2.6 & 0.3$\pm$0.1 & {\mybf 3.4$^{+1.7}_{-1.3}$} \\
00034169003 & 2015-11-27 20:15:59 & 2015-11-27 22:59:00 & 0.524--0.535 & 2.4 & 0.7$\pm$0.2 & {\mybf 9.2$^{+2.8}_{-2.3}$} \\
00034169004 & 2015-11-30 13:41:58 & 2015-11-30 16:17:56 & 0.789--0.799 & 2.4 & 0.2$\pm$0.1 & {\mybf 3.0$^{+1.8}_{-1.3}$} \\
00034169005 & 2015-12-03 11:58:57 & 2015-12-03 14:27:02 & 0.073--0.083 & 1.4 & {\mybf $<$0.52$^e$} & {\mybf $<$6.98$^e$} \\
00034169006 & 2015-12-08 09:58:58 & 2015-12-08 12:26:48 & 0.550--0.560 & 1.6 & {\mybf 0.8$^{+0.3}_{-0.2}$} & {\mybf 10.9$^{+3.7}_{-3.0}$} \\
00034169007 & 2016-01-09 07:47:58 & 2016-01-09 09:03:12 & 0.648--0.653 & 1.1 & {\mybf 1.0$^{+0.5}_{-0.4}$} & {\mybf 13.0$^{+6.3}_{-4.8}$} \\
00034169008 & 2016-01-11 04:35:58 & 2016-01-11 20:27:05 & 0.829--0.894 & 4.0 & 0.3$\pm$0.1 & {\mybf 4.6$^{+1.6}_{-1.3}$} \\
00034169009 & 2016-01-13 02:40:42 & 2016-01-13 08:20:50 & 0.016--0.039 & 4.1 & {\mybf $<$0.27$^e$} & {\mybf $<$3.62$^e$} \\
00034169010 & 2016-01-15 04:16:57 & 2016-01-15 14:35:20 & 0.216--0.258 & 4.7 & 0.5$\pm$0.1 & {\mybf 6.8$\pm$1.6} \\
00034169011 & 2016-01-17 07:13:58 & 2016-01-17 13:41:43 & 0.423--0.449 & 4.9 & 0.8$\pm$0.2 & {\mybf 10.5$\pm$2.0} \\
00034169012 & 2016-01-19 01:10:57 & 2016-01-19 19:35:58 & 0.592--0.667 & 3.6 & 0.5$\pm$0.1 & {\mybf 6.7$\pm$1.8} \\
\enddata
\tablecomments{
$^a$ {\mybf Phase zero} is defined {\mybf as} the epoch of maximum flux in the \textsl{Fermi} LAT. \\*
$^b$ The net exposure time spread over several snapshots.  Units are ks. \\*
$^c$ Count Rate is in the 2--10\,keV energy band.  Units are 10$^{-2}$\,counts s$^{-1}$. Errors are at the 1\,$\sigma$ level.\\* 
{\mybf $^d$ Unabsorbed \textsl{Swift} XRT flux in the 2.0--10.0\,keV bandpass converted with PIMMS.  Units are 10$^{-13}$\,\ergcm2s.\\*
$^e$ 3 $\sigma$ upper limits.
}
}
\label{table:xrt}
\end{deluxetable}

\begin{deluxetable}{lccccccc}
\tablecolumns{8}
\tabletypesize{\small}
\tablewidth{0pc}
\tablecaption{Australia Telescope Compact Array Radio Measurements}
\tablehead{
\colhead{Date} & \colhead{Mean MJD} & \colhead{Phase} &\colhead{Flux Density}
& \colhead{Flux Density} & \colhead{Error} & \colhead{Error}  \\
\colhead{} & \colhead{} & \colhead{} &\colhead{5.5 GHz (mJy)}
& \colhead{9 GHz (mJy)} & \colhead{5.5 GHz (mJy)} & \colhead{9 GHz (mJy)}
}
\startdata
2015-11-29 & 57355.58 & 0.693 & 0.440 & 1.160 & 0.093 & 0.133  \\
2015-12-03 & 57359.45 & 0.068 & -- & -- & 0.130 & 0.130 \\
2015-12-05 & 57361.81 & 0.298 & -- & -- & 0.180 & 0.130 \\
2015-12-08 & 57364.66 & 0.574 & 1.570 & 1.230 & 0.112 & 0.109 \\
2015-12-10 & 57366.76 & 0.778 & 1.020 & 1.040 & 0.095 & 0.104 \\
2015-12-25 & 57381.42 & 0.201 & -- & -- & 0.110 & 0.090 \\
2015-12-28 & 57384.63 & 0.513 & 1.560 & 1.240 & 0.205 & 0.118 \\
2016-01-02 & 57389.66 & 0.001 & -- & -- & 0.130 & 0.110 \\
2016-01-03 & 57390.55 & 0.088 & 0.790 & -- & 0.165 & 0.100 \\
2016-01-16 & 57403.40 & 0.335 & 1.010 & 0.670 & 0.078 & 0.078 \\
2016-01-17 & 57404.39 & 0.431 & 1.330 & 0.990 & 0.097 & 0.094 \\
2016-01-23 & 57410.38 & 0.013 & 0.520 & -- & 0.103 & 0.130 \\
2016-01-27 & 57414.35 & 0.398 & -- & -- & 0.120 & 0.100 \\
2016-01-31 & 57418.57 & 0.808 & 1.270 & 1.070 & 0.075 & 0.096 \\
2016-02-02 & 57420.37 & 0.982 & 1.040 & 0.580 & 0.087 & 0.085 \\
2016-03-02 & 57449.31 & 0.792 & 1.530 & 0.950 & 0.091 & 0.056 \\
2016-03-07 & 57455.02 & 0.346 & 1.040 & 0.660 & 0.104 & 0.087 \\
2016-03-12 & 57459.55 & 0.786 & 1.950 & 1.220 & 0.114 & 0.079 \\
\enddata
\tablecomments{The stated errors combine the statistical error, determined from RMS values in the region surrounding
\cxou, and a systematic error conservatively taken to be 5\% in the flux density scale between epochs.
}
\label{table:atca}
\end{deluxetable}


\begin{deluxetable}{lcc}
\tablecolumns{3}
\tablewidth{0pc}
\tablecaption{SOAR Radial Velocity Measurements}
\tablehead{
\colhead{Time} & \colhead {Phase} & \colhead{Velocity} \\
\colhead{(Barycentric Modified Julian Date)} &  & \colhead {(km s$^{-1}$)} }
\startdata
57366.0725460 & 0.711 & 284.9 $\pm$ 2.2 \\
57366.3190556 & 0.735 & 291.9 $\pm$ 2.2 \\
57378.1380143 & 0.883 & 289.6 $\pm$ 2.6 \\
57378.3045848 & 0.899 & 302.1 $\pm$ 2.6 \\
57383.0947451 & 0.364 & 300.5 $\pm$ 2.5 \\
57394.1203521 & 0.434 & 291.4 $\pm$ 2.3 \\
57417.0507903 & 0.660 & 288.5 $\pm$ 2.1 \\
57417.2847853 & 0.683 & 290.2 $\pm$ 2.2 \\
57425.0406091 & 0.436 & 286.6 $\pm$ 2.3 \\
57463.0104440 & 0.122 & 312.3 $\pm$ 2.2 \\
57473.0274778 & 0.094 & 312.8 $\pm$ 2.4 \\
57476.0021788 & 0.383 & 289.3 $\pm$ 2.3 \\
57494.9762214 & 0.225 & 293.1 $\pm$ 2.3 \\
\enddata
\tablecomments{Times are Barycentric Julian Dates (TDB) - 2400000.5}
\label{table:soar}
\end{deluxetable}


{\mybf
\begin{deluxetable}{ccccccc}
\tablecolumns{7}
\tabletypesize{\small}
\tablewidth{0pc}
\tablecaption{{\mybf Summary of X-ray Observations}}
\tablehead{
\colhead{{\mybf Mission/Instrument}} & \colhead{{\mybf Time}} & \colhead{{\mybf $\Gamma$}} & \colhead{{\mybf Flux}} & \colhead{{\mybf $L_{\rm x}$$^a$}} & \colhead{{\mybf Reference}} \\
\colhead{} & \colhead{{\mybf (MJD)}} & \colhead{} & \colhead{{\mybf (10$^{-13}$ \ergcm2s)}} & \colhead{{\mybf (10$^{35}$ \ergs)}} & \colhead{}}
\startdata
{\mybf \textsl{XMM-Newton}/EPIC} & {\mybf 53368} & {\mybf 1.57$^{+0.05}_{-0.06}$} & {\mybf 6.4$\pm$0.4$^b$} & {\mybf 2.32$\pm$0.14} & {\mybf \citet{Bamba2006}} \\
{\mybf \textsl{Chandra}/ACIS} & {\mybf 55599} & {\mybf 1.28$\pm$0.08} & {\mybf 3.71$\pm$0.10$^c$} & {\mybf 2.52$\pm$0.07} & {\mybf \citet{Seward2012}} \\
{\mybf \textsl{Chandra}/ACIS} & {\mybf 55600} & {\mybf 1.28$\pm$0.08} & {\mybf 4.70$\pm$0.12$^c$} & {\mybf 3.19$\pm$0.08} & {\mybf \citet{Seward2012}} \\
{\mybf \textsl{Swift}/XRT} & {\mybf 57347--57406} & {\mybf 1.3$\pm$0.3} & {\mybf $<$3.62--13.0$^d$} & {\mybf $<$1.09--3.9} & {\mybf Present work} \\
\enddata
\tablecomments{{\mybf Summary of the X-ray spectral parameters derived from the \textsl{XMM-Newton} \citep{Bamba2006}, \textsl{Chandra} \citep{Seward2012} and \textsl{Swift} observations (this work). \\*
$^a$ Luminosity converted to the 0.3--10.0\,keV bandpass with PIMMS assuming derived spectral parameters for
a distance of 50 kpc.\\*
$^b$ Absorbed \textsl{XMM-Newton} flux in the 0.5--10.0\,keV bandpass. \\*
$^c$ Absorbed \textsl{Chandra} flux in the 0.5--5.0\,keV bandpass. \\*
$^d$ Unabsorbed \textsl{Swift} XRT flux in the 0.3--10.0\,keV bandpass. Upper limits are 3 $\sigma$.}
}
\label{table:all_xray}
\end{deluxetable}
}

%% file: lmcp3_preprint.bbl
\begin{thebibliography}{}

\bibitem[Acero et al.(2015)]{Acero2015}
Acero, F., Ackermann, M., Ajello, M., et al.\ 2015, \apjs, 218, 23

\bibitem[Ackermann et al.(2016)]{Ackermann2016}
Ackermann, M., Albert, A., Atwood, W.~B., et al.\ 2016, \aap, 586, A71

\bibitem[Aliu et al.(2014)]{Aliu2014} 
Aliu, E., Archambault, S., Aune, T., et al.\ 2014, \apj, 780, 168

\bibitem[An et al.(2015)]{An2015}
An, H., Bellm, E., Bhalerao, V., et al.\ 2015, \apj, 806, 166

\bibitem[Atwood et al.(2009)]{Atwood2009} 
Atwood, W.~B., Abdo, A.~A., Ackermann, M., et al.\ 2009, \apj, 697, 1071 

\bibitem[Balucinska-Church \&  McCammon(1992)]{Balucinska1992}
Balucinska-Church, M., \& McCammon, D.\ 1992, \apj, 400, 699

\bibitem[Bamba et al.(2006)]{Bamba2006} 
Bamba, A., Ueno, M., Nakajima, H., Mori, K., \& Koyama, K.\ 2006, \aap, 450, 585 

\bibitem[Bodaghee et al.(2013)]{Bodaghee2013} Bodaghee, A., 
Tomsick, J.~A., Pottschmidt, K., et al.\ 2013, \apj, 775, 98 


\bibitem[Bozzetto et al.(2012)]{Bozzetto2012}
Bozzetto, L.~M., Filipovic, M.~D., Crawford, E.~J., De Horta, A.~Y., \& Stupar, M.\ 2012, Serbian Astronomical Journal, 184, 69

\bibitem[Burrows et al.(2005)]{Burrows2005}
Burrows, D.~N., Hill, J.~E., Nousek, J.~A., et al.\ 2005, \ssr, 120, 165 

\bibitem[Caliandro et al.(2015)]{Caliandro2015}
Caliandro, G.~A., Cheung, C.~C., Li, J., et al.\ 2015, \apj, 811, 68 

{\mybf
\bibitem[Capalbi et al.(2005)]{XRTGuide}
Capalbi, M., Perri, M., Saija, B., Tamburelli, F., \& Angelini, L.\ 2005,
http://swift.gsfc.nasa.gov/analysis/xrt\_swguide\_v1\_2.pdf
}

{\mybf
\bibitem[Cash(1979)]{Cash1979} Cash, W.\ 1979, \apj, 228, 939 
}

\bibitem[Clemens et al.(2004)]{Clemens2004}
Clemens, J.~C., Crain, J.~A., \& Anderson, R.\ 2004, \procspie, 5492, 331 

\bibitem[Coburn et al.(2002)]{Coburn2002}
Coburn, W., Heindl, W.~A., Rothschild, R.~E., et al.\ 2002, \apj, 580, 394

\bibitem[Corbel et al.(2012)]{Corbel2012} Corbel, S., Dubus, G., Tomsick, J.~A., et al.\ 2012, \mnras, 421, 2947 


\bibitem[Corbet et al.(2011)]{Corbet2011}
Corbet, R.~H.~D., Cheung, C.~C., Kerr, M., et al.\ 2011, The Astronomer's Telegram, 3221

\bibitem[Crampton et al.(1985)]{Crampton1985}
Crampton, D., Cowley, A.~P., Thompson, I.~B., \& Hutchings, J.~B.\ 1985, \aj, 90, 43 

{\mybf
\bibitem[de Grijs et al.(2014)]{deGrijs2014} 
de Grijs, R., Wicker, J.~E., \& Bono, G.\ 2014, \aj, 147, 122 
}

\bibitem[Dubus(2006)]{Dubus2006}
Dubus, G.\ 2006, \aap, 456, 801 

\bibitem[Dubus(2013)]{Dubus2013} Dubus, G.\ 2013, \aapr, 21, 64 

\bibitem[Dubus(2015)]{Dubus2015} Dubus, G.\ 2015, Comptes Rendus Physique, 16, 661 

\bibitem[Eastman et al.(2010)]{Eastman2010}
Eastman, J., Siverd, R., \& Gaudi, B.~S.\ 2010, \pasp, 122, 935

\bibitem[Evans et al.(2007)]{Evans2007}
Evans, P.~A., Beardmore, A.~P., Page, K.~L., et al.\ 2007, \aap, 469, 379

\bibitem[Faucher-Gigu{\`e}re \& Kaspi(2006)]{Faucher2006}
Faucher-Gigu{\`e}re, C.-A., \& Kaspi, V.~M.\ 2006, \apj, 643, 332 

\bibitem[{\nbf\Fermi} LAT Collaboration et al.(2009)]{FermiLAT2009}
{\nbf\Fermi} LAT Collaboration, Abdo, A.~A., Ackermann, M., et al.\ 2009, Science, 326, 1512 

\bibitem[{\nbf\Fermi} LAT Collaboration et al.(2012)]{FermiLAT2012}
{\nbf\Fermi} LAT Collaboration, Ackermann, M., Ajello, M., et al.\ 2012, Science, 335, 189

\bibitem[Foster (2012)]{Foster2012}
Foster, D. 2012, 
PhD Thesis Vanderbilt University

\bibitem[Hadasch et al.(2012)]{Hadasch2012}
Hadasch, D., Torres, D.~F., Tanaka, T., et al.\ 2012, \apj, 749, 54 

{\mybf
\bibitem[H.E.S.S.~Collaboration et al.(2015)]{HESS2015}
H.E.S.S.~Collaboration, Abramowski, A., Aharonian, F., et al.\ 2015, Science, 347, 406 
}

\bibitem[Hill et al.(2004)]{Hill2004}
Hill, J.~E., Burrows, D.~N., Nousek, J.~A., et al.\ 2004, \procspie, 5165, 217 

\bibitem[Kalberla et al.(2005)]{Leiden2005}
Kalberla, P.~M.~W., Burton, W.~B., Hartmann, D., et al.\ 2005, \aap, 440, 775

\bibitem[Kerr(2011)]{Kerr2011}
Kerr, M.\ 2011, \apj, 732, 38 

\bibitem[Liu et al.(2006)]{Liu2006}
Liu, Q.~Z., van Paradijs, J., \& van den Heuvel, E.~P.~J.\ 2006, \aap, 455, 1165 


\bibitem[Liu et al.(2007)]{Liu2007}
Liu, Q.~Z., van Paradijs, J., \& van den Heuvel, E.~P.~J.\ 2007, \aap, 469, 807 


{\mybf
\bibitem[Lyne et al.(2015)]{Lyne2015} Lyne, A.~G., Stappers, B.~W., Keith, M.~J., et al.\ 2015, \mnras, 451, 581 
}

{\mybf
\bibitem[Macri et al.(2006)]{Macri2006} Macri, L.~M., Stanek, K.~Z., Bersier, D., Greenhill, L.~J., 
\& Reid, M.~J.\ 2006, \apj, 652, 1133 
}

{\mybf
\bibitem[Malyshev \& Chernyakova(2016)]{Malyshev2016} Malyshev, D., \& Chernyakova, M.\ 2016, arXiv:1601.08216 
}


\bibitem[Marcote et al.(2015)]{Marcote2015}
Marcote, B., Rib{\'o}, M., Paredes, J.~M., \& Ishwara-Chandra, C.~H.\ 2015, \mnras, 451, 59 

\bibitem[Marcowith et al.(2016)]{Marcowith2016}
Marcowith, A., Bret, A., Bykov, A., et al.\ 2016, Reports on Progress in Physics, 79, 046901

\bibitem[Martins et al.(2005)]{Martins2005}
Martins, F., Schaerer, D., \& Hillier, D.~J.\ 2005, \aap, 436, 1049 

\bibitem[Mattox et al.(1996)]{Mattox1996}
Mattox, J.~R., Bertsch, D.~L., Chiang, J., et al.\ 1996, \apj, 461, 396 

\bibitem[Meurs \& van den Heuvel(1989)]{Meurs1989}
Meurs, E.~J.~A., \& van den Heuvel, E.~P.~J.\ 1989, \aap, 226, 88 

\bibitem[Mirabel \& Rodr{\'{\i}}guez(1998)]{Mirabel1998} Mirabel, I.~F., \& Rodr{\'{\i}}guez, L.~F.\ 1998, \nat, 392, 673 

{\mybf
\bibitem[Pietrzy{\'n}ski et al.(2013)]{Piet2013} Pietrzy{\'n}ski, G., 
Graczyk, D., Gieren, W., et al.\ 2013, \nat, 495, 76 
}

\bibitem[Sarty et al.(2011)]{Sarty2011}
Sarty, G.~E., Szalai, T., Kiss, L.~L., et al.\ 2011, \mnras, 411, 1293 

\bibitem[Sault et al.(1995)]{Sault1995}
Sault, R.~J., Teuben, P.~J., \& Wright, M.~C.~H.\ 1995, Astronomical Data Analysis Software and Systems IV, 77, 433
 
\bibitem[Scargle(1982)]{Scargle1982}
Scargle, J.~D.\ 1982, \apj, 263, 835

\bibitem[Seward et al.(2012)]{Seward2012} 
Seward, F.~D., Charles, P.~A., Foster, D.~L., et al.\ 2012, \apj, 759, 123

\bibitem[Spitkovsky(2006)]{Spitkovsky2006}
Spitkovsky, A.\ 2006, \apjl, 648, L51 

\bibitem[Strader et al.(2015)]{Strader2015}
Strader, J., Chomiuk, L., Cheung, C.~C., Salinas, R., \& Peacock, M.\ 2015, \apjl, 813, L26

\bibitem[Tavani et al.(2009)]{Tavani2009} Tavani, M., Bulgarelli, A., Piano, G., et al.\ 2009, \nat, 462, 620 


\bibitem[Udalski et al.(2015)]{Udalski2015}
Udalski, A., Szyma{\'n}ski, M.~K., \& Szyma{\'n}ski, G.\ 2015, \actaa, 65, 1

\bibitem[van der Marel et al.(2002)]{vanderMarel2002}
van der Marel, R.~P., Alves, D.~R., Hardy, E., \& Suntzeff, N.~B.\ 2002, \aj, 124, 2639 

\bibitem[Wilms et al.(2000)]{Wilms2000}
Wilms, J., Allen, A., \& McCray, R.\ 2000, \apj, 542, 914

\bibitem[Wilson et al.(2011)]{Wilson2011}
Wilson, W.~E., Ferris, R.~H., Axtens, P., et al.\ 2011, \mnras, 416, 832 

\end{thebibliography}
